\documentclass[10pt,twocolumn]{article}

\usepackage[letterpaper,margin=0.75in]{geometry}
\usepackage{amsmath,amssymb}
\usepackage{newtxtext,newtxmath}
\usepackage[T1]{fontenc}
\usepackage[utf8]{inputenc}
\usepackage{booktabs}
\usepackage{url}
\usepackage{listings}
\usepackage{enumitem}
\usepackage{authblk}
\usepackage{xurl}

\usepackage{hyperref}

\usepackage{xcolor}
\definecolor{linkblue}{HTML}{1A5FB4}
\hypersetup{
  colorlinks=true,
  linkcolor=linkblue,
  citecolor=linkblue,
  urlcolor=linkblue,
  breaklinks=true,
}
\setlist{nosep,leftmargin=*}

\newcommand{\code}[1]{\texttt{\small #1}}
\newcommand{\ie}{i.e.\ }

\title{\bf Git Hash Chain Malleability}

\author{Jacob Ginesin}
\affil{Carnegie Mellon University \quad \texttt{ginesin@cmu.edu}}

\date{}

\begin{document}


\twocolumn[
  \begin{@twocolumnfalse}
  \maketitle
  \begin{abstract}
  \noindent

  Git commit signing is widely entrusted to serve as evidence that 
  a commit hash uniquely and immutably identifies a specific piece of signed content.
  We show this invariant does not hold. Given any signed commit, an attacker 
  \textit{without access to the signing key, and without breaking SHA2}, can produce a second, distinct commit 
  with an identical tree, identical metadata, a valid signature, and a ``Verified'' 
  badge from a Git Forge such as Github, differing only in its commit hash.
  The modified commit cascades to modify the values of each subsequent, dependent commit hashes; hence we introduce the terminology ``hash chain malleability'' to describe this phenomenon. 
  The malleability in signed Git hashes is feasible due to the inherent malleability present in many of the 
  data representations that constitute a commit. In this paper we show
  three such malleation routes: (i) algebraic inversion $s \mapsto n-s$ for ECDSA; 
  (ii) structural insertion of an unhashed OpenPGP subpacket (RFC~4880 \S5.2.3) for
  RSA and EdDSA; and (iii) non-canonical DER length re-encoding (X.690 \S10.1)
  inside the CMS envelope for S/MIME. Algebraic inversion for ECDSA signatures and
  subpacket insertion were found to pass local verification (\code{git verify-commit}),
  and all three methods yield a persistent, independent ``Verified'' record on Github. 
  We discuss the consequences of Git hash chain malleation 
  for hash-based commit blocking, dependency pinning
  (Nixpkgs, Go modules, Github Actions), and reproducible-build systems that
  treat the commit hash as a content-addressable primary key, and we provide
  proof-of-concept tooling that automates all three routes.
  \end{abstract}
  \vspace{1.5em}
  \end{@twocolumnfalse}
]

\section{Introduction}

Git identifies every object by the hash of its serialized content.
For a commit, this hash covers the tree, the parent list, author, and committer metadata,
the commit message, and---crucially---the raw bytes of any embedded signature. 
Because the hash is a deterministic function of these bytes, the commit hash is routinely
relied upon as a globally unique, immutable, content-addressable identifier: 
CI pipelines pin to them, dependency managers lock to them, and incident-response tooling 
blocks or reverts by them. Quoting Git's own documentation:

\begin{quote}
  ``Using a cryptographically secure hash function brings additional advantages: 
    Object names can be signed and third parties can trust the hash to address the signed
    object and all objects it references.'' --- Git's \code{hash-function-transition} documentation \cite{git-hash-function-transition}
\end{quote}

Commit signing is intended to strengthen the trust placed into the subject commit hash. A signature over a commit payload intends to act as a verifiable, unforgeable method for tracking contribution provenance; therefore, projects or ecosystems sensitive to supply-chain security vulnerabilities oftentimes enforce its adoption. Indeed, a post-mortem of the recent \emph{trivy-action} supply-chain compromise showed that malicious commits uploaded in-place of the original commits were distinguished by their \emph{signature validity}~\cite{burckhardt2026trivy}, directly demonstrating the utility and security benefit of signing commits.

In this note, we show that signature validity and hash uniqueness interact in a way their consumers do not necessarily expect. Digital signature schemes oftentimes admit \emph{malleability}: given a valid signature, one can often derive a second, syntactically distinct signature that verifies against the same message and key without knowledge of the private key. Applied to a signed commit,
malleation changes the signature bytes, and hence the commit hash, while preserving
both the content and the verification result. This directly contradicts the claimed uniqueness of hash-based identifiers for Git objects:
\begin{quote}
  ``\texttt{object}: 
  The unit of storage in Git. It is uniquely identified by the SHA-1 of its contents. Consequently, an object cannot be changed.'' --- Git's \texttt{gitglossary} documentation \cite{gitglossary}
\end{quote}

The malleated commit cascades to malleate the values of all the subsequent, dependent commit hashes. Hence, we introduce the terminology to describe this phenomenon: ``hash chain malleability.''

In this note, we make the following contributions:
\begin{itemize}
  \item We characterize three signature malleation techniques applicable to commit signing: mathematical malleation, structural malleation, and encoding malleation. Together, these three techniques enable the successful malleation of commit signatures for every GPG-based signature scheme Github supports, including ECDSA, RSA, EdDSA, and S/MIME/CMS. 
  \item We observe that these are not merely local parser quirks: Github's server-side verifier accepts malleation for all of the aforementioned signature schemes, and its persistent verification record~\cite{ghpersist}, keyed on commit hash, produces multiple independent
  ``Verified'' entries for the same content (\S\ref{sec:github}).
  \item We analyze the impact on hash-based blocking, dependency pinning, and
  reproducible-build verification, where the commit hash is used as a content
  primary key (\S\ref{sec:impact}).
  \item We provide proof-of-concept tooling automating all three malleation techniques over an arbitrary commit graph, including graph-consistent rewriting of descendant commits (\S\ref{sec:poc}).
\end{itemize}
All of our code and proof-of-concepts are publicly available and readily reproducible via a Nix Flake: \url{https://github.com/JakeGinesin/git-chain-malleator}.

\section{Background}
\label{sec:background}

A Git commit is a plaintext object whose object name is the hash of its full serialization. That serialization comprises the tree, the ordered parent list, the author and committer lines, the commit message, and, when the commit is signed, a \code{gpgsig} header carrying the signature. The signed payload is precisely the commit object with the \code{gpgsig} header removed; the signature is thus computed over everything \emph{except} itself, and is then folded back into the object that the hash covers.

Two consequences follow, and together they are the precondition for every technique in this note. First, because the signature bytes lie inside the hashed region, any change to those bytes changes the commit hash while leaving the tree, parents, and message untouched. Second, and less obviously, verification constrains a signature's \emph{semantic} content but not its \emph{exact byte encoding}: a verifier checks that some valid signature over the payload is present, not that it is the unique admissible encoding of one. The gap between ``a valid signature'' and ``this signature'' is the room in which a malleator operates.

We stress that this is orthogonal to the collision resistance of the commit hash function, whether the SHA-1 in use today or the SHA-256 (\ie the SHA-2 family) that Git's ongoing transition targets~\cite{git-hash-function-transition}. We do not exhibit two distinct contents sharing one hash. We exhibit one logical content admitting many byte-distinct but semantically identical serializations, each carrying a valid signature and each hashing to a \emph{different} value. The malleability we describe therefore survives the SHA-256 migration untouched.

\section{Malleation Techniques}
\label{sec:techniques}

We organize the malleation of a commit signature into three techniques, distinguished by the layer at which the byte change is introduced. \emph{Mathematical malleation} alters the signature value itself, exploiting an algebraic symmetry of the signature scheme, and yields a value that is still a valid signature under the same key. \emph{Structural malleation} leaves the signature value intact and edits the surrounding container in a region the signature does not cover. \emph{Encoding malleation} leaves both the signature value and the container's logical content intact, and re-serializes some field into an equivalent but non-canonical byte encoding. Which technique applies is a function of the signature scheme, as detailed below and summarized in Table~\ref{tab:results}.

\subsection{Mathematical malleation (ECDSA)}

The malleability of ECDSA is a classical property of the scheme~\cite{x962} and is familiar from the Bitcoin transaction-malleability literature~\cite{bip62}. For a curve of order $n$ and a signature $(r, s)$, the pair $(r,\, n - s)$ is an equally valid signature over the same message: verification recovers a curve point and tests only its $x$-coordinate, and $s$ and $n - s$ yield points reflected across the $x$-axis that therefore share it. The transformation consumes only the public parameter $n$; the private key is never involved.

We recover $s$ from the OpenPGP multi-precision integer in the \code{gpgsig} header, infer the curve from the bit length of the scalar (thereby mapping it to the order of $P$-256, $P$-384, or $P$-521) and re-emit the integer for $n - s$. Because the re-encoded scalar may occupy a different number of octets than the original, the enclosing packet's length header is recomputed. As the resulting signature is genuinely valid, it is accepted not only by Github but by strict local \code{git verify-commit}.

\subsection{Structural malleation (RSA and EdDSA)}

RSA is not malleable in the algebraic sense, and EdDSA is designed to preclude scheme-level malleability altogether. For these we leave the signature value untouched and target the OpenPGP container instead. A version-4 signature packet carries two subpacket regions: a \emph{hashed} region, which is covered by the signature, and an \emph{unhashed} region, which per RFC~4880 \S5.2.3 ``is not cryptographically protected by the signature and should include only advisory information''~\cite{rfc4880}. A conforming verifier folds only the packet body through the end of the hashed subpackets into its digest; the unhashed region is excluded by construction.

We therefore append a single well-formed subpacket to the unhashed region: a non-critical subpacket of private/experimental type~100, which \S5.2.3.1 requires a conforming implementation to ignore.
We then adjust the unhashed-length field accordingly, re-emitting the outer packet header in long-length form since the body has grown. Because the edit falls entirely outside the signed region, the digest the signature attests is unchanged and the signature remains valid. Consequently the malleated commit is accepted by \emph{both} Github and strict local \code{git verify-commit}. Curiously, the property RFC4880 relies on to permit advisory metadata (i.e. that unhashed subpackets are not signed) is precisely what lets the commit hash be rewritten while preserving verification.

\subsection{Encoding malleation (S/MIME)}

An S/MIME commit signature is a CMS \code{SignedData} structure~\cite{rfc5652}, encoded in ASN.1 and embedded in the \code{gpgsig} header. The encoding is required to be DER, the distinguished ``canonical'' subset of BER: X.690 \S10.1 enforces that every length field use the shortest admissible form~\cite{x690}. A length of $32$ must be written as the single octet \code{0x20}, and \emph{not} as the two-octet BER long form \code{0x81\,0x20}, even though a BER decoder accepts both as equal.

We select a length field on an interior element of the CMS envelope that is specifically not a value other than the outermost \code{SEQUENCE}, and rewrite it from canonical short form into the equivalent long form, propagating the resulting size delta outward through each enclosing length field. The blob ceases to be valid DER but remains valid BER; neither the signed content nor the signature value is touched, so the signature still verifies. Because the manipulation lives in the envelope encoding rather than in the signature, it is \emph{algorithm-agnostic}: it succeeds regardless of whether the S/MIME key is RSA or ECDSA, a fact we confirmed against RSA-2048, RSA-4096, ECDSA $P$-256, and ECDSA $P$-384 signers.

Here local and server behavior diverge. A strict CMS parser that enforces DER (e.g. libksba, and therefore \code{gpgsm}) rejects the non-canonical length, so local verification fails; Github's more permissive parser accepts it and marks the commit ``Verified.'' The same object thus draws opposite verdicts from two verifiers both nominally implementing CMS.

\begin{table}[t]
\centering
\small
\begin{tabular}{@{}llcc@{}}
\toprule
Scheme & Technique & Local & Github \\
\midrule
ECDSA   & mathematical ($s\!\to\!n\!-\!s$) & accept & accept \\
RSA     & structural (subpacket)          & accept & accept \\
EdDSA   & structural (subpacket)          & accept & accept \\
S/MIME  & encoding (DER length)           & reject & accept \\
\bottomrule
\end{tabular}
\caption{Malleation technique per scheme, and the verdict of strict local
verification (\code{git verify-commit} for OpenPGP, \code{gpgsm} for S/MIME)
versus Github's server-side verifier. Every technique preserves the tree and
yields a distinct commit hash. Only S/MIME's encoding malleation is caught
locally; the three OpenPGP routes survive even strict local verification. The results are up-to-date for Github at the time of writing.}
\label{tab:results}
\end{table}

\section{Server-Side Acceptance and Record Duplication}
\label{sec:github}

What elevates these techniques from local curiosities to a supply-chain concern is that Github's server-side verifier accepts all of them. Github does not canonicalize a signature container before verifying it: it does not enforce DER strictness on CMS envelopes, does not strip or normalize OpenPGP unhashed subpackets, and accepts non-canonical ECDSA scalars. Each accepted commit is then issued an independent verification record, which Github persists and keys on the commit hash; by design this record is never re-evaluated, so that a commit retains its ``Verified'' status even after the signing key is rotated or revoked~\cite{ghpersist}.

A single signed logical commit can be represented by an unbounded family of byte-distinct hashes, each of which Github independently and durably marks ``Verified,'' with nothing in its interface or API to indicate that they encode identical content. Pushed to two branches, an original and its malleated twin are presented by Github's comparison view as unrelated history ``1 commit ahead and 1 commit behind''despite byte-identical trees and metadata. The invariant that a signed, verified commit hash is a unique handle for a specific piece of signed content therefore does not hold on Github, and by the same methodology does not hold on any Git forge that verifies signatures without first canonicalizing their encoding.

\section{Potential Impact}
\label{sec:impact}

The direct effect of the techniques presented in this paper is narrower than either a
hash collision or a content forgery. Given a signed commit, an attacker can
derive a second, validly signed serialization of the same commit payload and
therefore a second Git object identifier. The tree, parent list, author and
committer fields, and commit message of the malleated commit remain unchanged.
Exploitation nevertheless requires a channel through which the attacker can
publish or serve the alternate object and induce a consumer to trust it, such
as a writable repository, fork, mirror, bundle, or source-distribution service.
A consumer already anchored to the original full commit hash remains anchored
to the original object.

\paragraph{Limits on propagation through signed history.}

A malleated commit anywhere in the hash chain does not automatically produce a fully verifying alternate history -- only the malleation of the \textit{most recent} signed commit preserved validity.
This is because the malleation of a commit requires recalculating and replacing all 
child commit hashes to maintain correctness, in-turn invalidating all signatures on commits downstream from the malleated commit. 
In the general case, the result is a validly signed twin of one commit and a point of
diverging history, rather than a replacement chain in which every descendant
retains its original valid signature.

\paragraph{Bypass of hash-only blocking and deduplication.}
Incident-response and push-protection tooling that blocks or reverts a specific malicious commit \textit{specifically by commit hash} may be circumvented silently. A block targets $\mathrm{hash}(\mathrm{original})$; the attacker re-pushes a byte-distinct but content-identical and still-valid signed commit under $\mathrm{hash}(\mathrm{ghost}) \neq \mathrm{hash}(\mathrm{original})$, which is not on the blocklist. No metadata or tree change is required, and the substitution leaves no trace in the commit history.

\paragraph{Repository equivocation and mirror divergence.}
A hostile mirror or fork can publish a malleated Git hash chain alongside, 
or instead of,
the canonical Git hash chain while presenting the same checked-out tree and a valid
signature on the malleated commit. Where unsigned descendants can also be
rewritten, the mirror may expose a longer alternate graph whose object IDs
diverge from the canonical graph beginning at the malleated commit. Repository
comparison, synchronization, and incident-response systems may consequently
report a substantive history divergence even when the relevant source trees
are equivalent.

This is best understood as an equivocation or reconciliation problem rather
than an undetectable content substitution. A consumer that knows and checks
the expected canonical tip of the Git hash chain 
remains protected, whereas a consumer that accepts
any commit carrying an approved signer identity or a forge-provided
``Verified'' status may accept either representation.

\paragraph{Dependency pinning and lockfile ambiguity.}
Hash-chain malleability does \emph{not} replace different content beneath an
existing commit hash pin of dependencies. A Github Actions reference pinned to a full SHA,
a Go pseudo-version identifying a particular revision, or another dependency
record containing the original commit ID continues to request that original
object. An attacker may produce a different identifier rather than a second
object with the pinned identifier. Full-SHA pinning therefore retains its
principal integrity property~\cite{andrewlock-provenance}.

The impact is instead the creation of aliases at the signed-content level.
A lockfile regenerated from an attacker-controlled mirror or mutable reference
may record the ghost hash even though the fetched tree is unchanged. Systems
may then produce duplicate cache entries, unnecessary lockfile changes, or
inconsistent allowlist, denylist, review, and provenance records for equivalent
source. Nix-style schemes that pair a revision with an independently checked
content hash, such as a flake \code{narHash} or fixed-output hash, retain their
content-integrity check~\cite{nixdev-pinning}.




\paragraph{Reproducible-build and provenance verification.}
Systems that adopt the Git hash as the basis of truth for reproducible builds or
artifact provenance inherit the weakness presented by hash chain malleability
wholesale: a signed, ``Verified'' hash is
no longer a unique commitment to content, and any downstream check deriving trust
from it is correspondingly weakened. The SLSA
Source Track (v1.2) treats the commit hash as the immutable revision identifier
from Source Level~1 onward~\cite{slsa-source-req,safeguard-source}, under a threat
model in which the adversary creates a repository revision and induces consumers to
fetch it~\cite{slsa-source-verify}, precisely the capability our malleation grants.
The same commit is the ground truth in the build track, whose provenance records the
resolved source digest~\cite{slsa-provenance} and whose verifier consults that SHA
for policy decisions~\cite{slsa-verifier}. Keyless commit signing raises the stakes
further by turning the hash into a durable, publicly logged trust anchor: Sigstore's
Gitsign, building on the transparency model of Newman et~al.~\cite{sigstore-ccs},
writes a hashed record over the commit SHA into the append-only Rekor log, keeping
commits verifiable by hash after the ephemeral certificate
expires~\cite{sigstore-gitsign}. Because our second commit carries a valid signature
and earns its own independent log entry, this anchor is duplicated rather than
defended. Dependency managers make the commit the addressing primitive directly, as
a Nixpkgs fetcher revision~\cite{nixdev-pinning}, a Go module pseudo-version, or a
pinned Github Actions SHA adopted specifically to resist supply-chain
tampering~\cite{andrewlock-provenance}, while reproducible builds compound the
exposure by feeding the commit, often as the build timestamp, into the produced
artifact~\cite{reproduciblebuilds-versioninfo,haven-repro}. The exposure is uneven:
pinning schemes that pair the commit with an independent content hash of the fetched
tree, such as Nix fixed-output derivations~\cite{nixdev-pinning}, retain a residual
check, whereas verifiers that treat \code{git verify-commit} as authoritative are
fully exposed, since that command does not validate certificate claims and accepts
our malleated commits~\cite{kenmuse-gitsign}.

\paragraph{Bypass of release immutability.}
An attacker or maintainer may mint multiple content-identical commits and publish
each as its own immutable release. Each receives an independent attestation and
a ``Verified'' badge, so immutability's tag $\rightarrow$ hash $\rightarrow$ asset
binding does not prevent an unbounded number of equally verified releases for identical content.

\section{Proof of Concept}
\label{sec:poc}

We provide tooling that automates all three techniques.\footnote{\url{https://github.com/JakeGinesin/git-chain-malleator}} \code{interactive-malleator.py} reads a commit's \code{gpgsig} header, identifies the public-key algorithm, and dispatches the corresponding technique: mathematical inversion for ECDSA, and unhashed-subpacket insertion for RSA and EdDSA. Because malleating a commit changes its hash, and each descendant names that hash in its \code{parent} field, the tool then rewrites the descendant chain such that the graph remains well-formed, and advances the branch pointer to the new head. \code{setup-ghost.py} builds a repository containing an S/MIME-signed commit from a \code{.p12} certificate and derives a ``ghost'' twin via non-canonical DER length re-encoding.

A per-scheme test suite asserts, for each technique, that the malleated hash differs from the original and that the tree is preserved; for the OpenPGP routes it further asserts that strict local \code{git verify-commit} returns the verdict of Table~\ref{tab:results}, and for S/MIME that the byte-level difference is confined to the \code{gpgsig} header.

Besides the results on a local \code{git verify-commit}, we additionally provide Github repositories where an x509 S/MIME-signed commit\footnote{\url{https://github.com/JakeGinesin/github-x509-malleability}} and an GPG EdDSA-signed commit\footnote{\url{https://github.com/JakeGinesin/github-gpg-malleability}} were malleated using our tooling. 

\section{Discussion and Mitigations}

The three techniques have distinct root causes but share a single theme: verification is more permissive than the identity model built atop it assumes. The corresponding fixes are equally distinct. For S/MIME, a Git forge should enforce DER canonicalization per X.690 \S10 before accepting a CMS envelope, rejecting or normalizing any non-canonical BER. For OpenPGP, a verifier should strip or ignore unhashed subpackets before deriving any identity signal from a signature, as recommended for abuse-resistant OpenPGP handling~\cite{dkg}. For ECDSA, canonicalizing to a low-$s$ representation, as Bitcoin adopted via BIP-62~\cite{bip62}, closes the algebraic route; but because ECDSA malleation already passes strict local verification, this must be enforced by every consumer that relies on hash uniqueness rather than assumed to hold at signing time. 

More broadly, formally verified parsers such as those generated by Vest~\cite{vest} guarantee round-trip non-malleability by construction. The CMS/DER parsers in Git forges that ingest S/MIME envelopes, the OpenPGP packet parsers that separate hashed from unhashed subpackets, and the DER-based ECDSA signature decoders that consumers rely on for hash uniqueness would all benefit from the adoption formally verified non-malleable parsing techniques, implicitly presenting the class of issues we presented.

In general, no consumer should treat a signed, verified commit hash as a unique handle for content. Hash-based deduplication, blocking, and pinning belong \emph{downstream} of verification and content canonicalization rather than on the raw hash of a signed object whose encoding an adversary controls. 

\section{Conclusion}

We have shown that signed Git commits are malleable across every signature scheme Github GPG verifies, that three techniques---mathematical, structural, and encoding-based---suffice to cover them, and that Github accepts all three and durably records each malleated commit as ``Verified'' under a distinct hash. None of this disturbs the collision resistance of the underlying hash function; the attack instead breaks the assumption that a verified hash is a \emph{unique} name for signed content. The practical consequence reduces to a single sentence: a signed commit, nor a ``Verified'' badge on a commit hash from Github can serve as a unique identifier for a piece of content; and yet, systems treat it as such. Hash-based blocking, dependency pinning, and reproducible-build verification among them should canonicalize and verify before, not after, they place trust in the hash.

\section{Acknowledgements}
We acknowledge the helpful discussions and feedback from William Enck, Ezri Zhu, and Edward Berman.

\section{Ethics and Responsible Disclosure}
We complied with standard responsible disclosure procedure throughout this work. 
We initiated responsible disclosure with GNU (for Libksba) and Git on Jan 16th 2026, as well as Github on March 3rd 2026;
as of writing, unfortunately this issue has failed to be addressed by Git nor any Git forge.

\bibliographystyle{plain}
\bibliography{refs}

\end{document}